\documentclass[twocolumn,showpacs,amsmath,amssymb]{revtex4}

\usepackage{graphicx}% Include figure files
\usepackage{dcolumn}% Align table columns on decimal point
\usepackage{bm}% bold math

\begin{document}

\title{Onset of mechanical stability in random packings of frictional spheres}

\author{Melissa Jerkins} \author{Matthias Schr\"oter}
\email{matthias.schroeter@ds.mpg.de} \author{Harry L. Swinney}
\email{swinney@chaos.utexas.edu}
\affiliation{Center for Nonlinear Dynamics and Department of Physics\\
University of Texas at Austin, Austin, Texas 78712}

\author{Tim J. Senden} \author{Mohammad Saadatfar} \author{Tomaso
Aste} \affiliation {Department of Applied Mathematics, Research School
of Physical Sciences and Engineering, The Australian National
University, 0200 Canberra, ACT, Australia}

\date{\today}

\begin{abstract}

Using sedimentation to obtain precisely controlled packings of
noncohesive spheres, we find that the volume fraction $\phi_{\rm RLP}$
of the loosest mechanically stable packing is in an
operational sense well defined by a limit process. This random
loose packing volume fraction decreases with decreasing pressure $p$
and increasing interparticle friction coefficient $\mu$.  Using X-ray
tomography to correct for a container boundary effect that depends on
particle size, we find for rough particles in the limit $p \rightarrow
0$ a new lower bound, $\phi_{\rm RLP} = 0.550 \pm 0.001$.

\end{abstract}


\maketitle

{\it Introduction. --} If granular materials such as sand, sugar, or
snow are excited strongly (e.g. by shaking or shearing), they exhibit
fluid-like behavior. However, after the excitation stops, dissipation
quickly produces a static packing that is mechanically stable under
its own weight. Experiments
\cite{scott:60,rutgers:62,onoda:90,ojha:00,valverde:04,valverde:06,dong:06,umbanhowar:06}
and simulations
\cite{makse:00,ohern:02,zhang:05,shundyak:07,silbert:08} have shown
that the volume fraction has a well defined lower limit, $\phi_{\rm RLP}$,
called Random Loose Packing (RLP).

The value of $\phi_{\rm RLP}$ depends on the particle-particle
interactions.  Packings of cohesive particles like fine powders are
stable under their own weight for values of $\phi_{\rm RLP}$ as low as
0.15 \cite{valverde:04,valverde:06,dong:06,umbanhowar:06}. However,
many granular materials do not exhibit cohesive forces.  Simulations
of frictionless elastic noncohesive spheres have found the onset of a
finite bulk modulus at the jamming point, $\phi_{\rm J} \approx 0.64$
\cite{makse:00,ohern:02,zhang:05}. %ohern:03
Real spheres have friction and then it has been suggested that $\phi_{\rm RLP}$ 
depends on the density difference
between the particles and the surrounding fluid
\cite{scott:60,rutgers:62,onoda:90,ojha:00}. The lowest volume
fraction reported thus far, $\phi_{\rm RLP} = 0.555$, was observed for
slowly sedimenting spheres in a liquid of nearly the same density
\cite{onoda:90}.

Here we demonstrate a limit process that yields well defined values of
$\phi_{\rm RLP}$ that depend on  pressure  and 
coefficient of friction. The results are discussed in the context of a
statistical mechanics approach based on the ensemble of all
mechanically stable configurations \cite{edwards:07}.

%%%%%%%%%%%%%%%%%%%%%%%%%%%%%%%%%%%%%%%%%%%%%%%%%%%%%%%%%%%%%%%%%%%%%%%%%%%%%%%%%%%%%%%%%%%%%%%%%%%%%%%%%%%%%%%%%%%%%

\begin{figure}
  \begin{center}
    \includegraphics[height=88 mm,angle=-90]{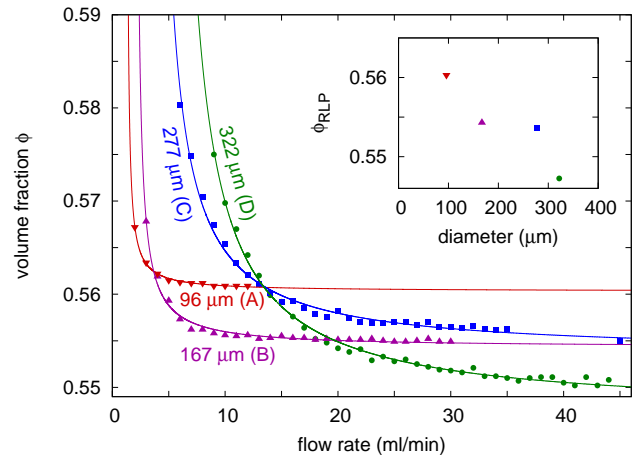}
    \caption{(Color online) The approach to Random Loose Packing in a
    limit process is achieved using flow pulses in a liquid fluidized bed. Data
    for different particle diameters were fit to (\ref{eq:rlp_fit}) to obtain $\phi_{RLP}$.
    Letters in parentheses refer to the particle samples in Table
    \ref{tab:particles}.  Inset: Diameter dependence of 
    $\phi_{\rm RLP}$ without container size correction (see text).  Sample height was 97 mm at RLP;
    five flow pulses were averaged for each flow rate.  
}
    \label{fig:diameter}
  \end{center}
\end{figure}

{\it Experiment. --} Mechanically stable packings of glass spheres
were prepared by allowing the particles to sediment following flow
pulses in a water fluidized bed. The fluidized bed was contained in a
vertical polycarbonate tube with an inner diameter $D$ of 12.8 mm and
a length of 230 mm. The tube's bottom end was closed by a distributor
consisting of a porous bronze disc (height, 8 mm; nominal pore size,
25 $\mu$m). A programmable syringe pump (Harvard Apparatus) created
pulses of constant flow rate $Q$. During a flow pulse of 2 minutes
length the granular medium fluidized and expanded until it reached a
steady state height. After each flow pulse, the particles sedimented
onto the distributor and formed a mechanically stable packing whose
volume fraction depended on $Q$, as shown in
Fig.~\ref{fig:diameter}. A higher value of $Q$ resulted in a more
expanded fluidized bed, longer sedimentation time, and lower $\phi$ of
the sediment.  Packings created in this way are independent of the
state of the sample prior to the last flow pulse \cite{schroeter:05}.
This property is important for any statistical mechanics approach
\cite{aste:07b}.  The volume fractions in Fig.~\ref{fig:diameter} are
averaged over the whole sample: $\phi = m / \rho A h_{\rm sed}$, where
$\rho$ is the particle density, $m$ is the total mass of all the
spheres, $A$ is the cross sectional area of the tube, and $h_{\rm
sed}$ is the height of the sedimented sample determined from images.

The properties of the different samples of particles are given in
Table \ref{tab:particles}. The density $\rho$ of the particles in each
sample was measured with an accuracy of 0.06 \% %$\pm$ 2 mg/cm$^3$
using a Gay-Lussac specific gravity bottle
%and additionally for the 250 and 265  $\mu$m with
and a Micromeritics AccuPyc 1330 gas pycnometer; the average $\rho$
was 2.48 g/cm$^3$.  To characterize the frictional properties of the
samples we measured the angle of repose under water: a beaker
containing a layer of particles about 5 mm high was tilted until the
particles started to move. To obtain an especially rough sample (F) we
soaked part of sample C for 3 hours in hydrofluoric acid.  Sample E
consists of spheres that were smoothed by exposure to more than 45,000
flow pulses in a fluidized bed \cite{schroeter:05}. 

\begin{table}
\caption{Properties of the different samples of glass
spheres. Particle diameters $d$ and standard deviations $\sigma$ were
measured with a Camsizer (Retsch Technology).  Angles of repose under water were
averaged over 10 measurements.}
\label{tab:particles}
\begin{tabular}{c|c|c|c|c}
sample & $d$ ($\mu$m) & $\sigma$(\%)  & supplier & angle of repose  \\ % MW3 and Sigma3
\hline
A & 96  & 15.6 & Cataphote & 24.8 $\pm$ 1.0  \\
B & 167 & 16.1  & Cataphote & 26.1 $\pm$ 0.7  \\
C & 277 & 7.6 & Cataphote & 25.3 $\pm$ 0.8  \\
D & 322 & 9.3 & Cataphote & 25.5 $\pm$ 0.7  \\
E & 261 & 5.0  & MoSci     & 24.0   $\pm$ 0.8  \\
F & 257 & 7.8  & Cataphote     & 27.7 $\pm$ 1.3  \\
G & 255 & 2.7 & MoSci     &   26.6 $\pm$ 0.7     \\
\end{tabular}
\end{table}

%%%%%%%%%%%%%%%%%%%%%%%%%%%%%%%%%%%%%%%%%%%%%%%%%%%%%%%%%%%%%%%%%%%%%%%%%%%%%%%%%%%%%%%%%%%%%%%%%%%%%%%%%%%

{\it RLP is defined by a limit process. --} The main improvement over
earlier studies using sedimenting particles \cite{onoda:90,dong:06} is
that our control of $Q$ allows us to change the sedimentation time
independent of the liquid density. This procedure reveals the
convergence of $\phi$ to $\phi_{\rm RLP}$.  The observation that the
slowest relaxing preparation yields the loosest packings agrees with
simulations of frictional discs and spheres
\cite{zhang:05,shundyak:07}.  Figure~\ref{fig:diameter} shows that
$\phi (Q)$ is well described by the fit function used in
\cite{schroeter:05},
\begin{equation}
\label{eq:rlp_fit} 
\phi (Q) = \phi_{\rm RLP} + \frac{a}{Q-b},
\end{equation}
which we use to determine $\phi_{\rm RLP}$.

\begin{figure*}
  \begin{center}
    \includegraphics[width=17cm]{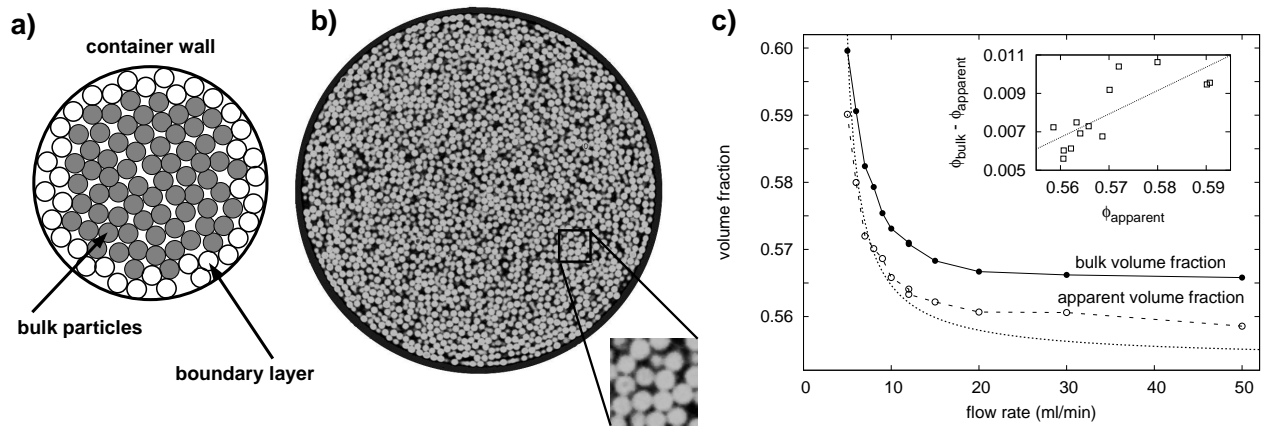}
    \caption{(a) Illustration of the necessity of a finite size
    correction of the volume fraction determined from the total sample
    volume. Particles near the boundary have a lower volume fraction
    than those in the bulk. 
    (b) Cross section of a 3-dimensional x-ray tomogram of the
    fluidized bed; the particles have a diameter of 255 $\mu$m (sample G).
    The inner diameter of the polycarbonate tube (black circle) is
    12.8 mm.  (c) Bulk volume fraction
    for particles that are at least $4d$ away from
    boundary, and the apparent volume fraction $\phi_{\rm apparent}$ for
    all particles.  The dotted line
    corresponds to the fit to the 277 $\mu$m particles in
    Fig.~\ref{fig:diameter}. 
    The inset shows a linear fit  (\ref{eq:finite_size_corr}) to the
    difference between bulk and apparent volume fractions.
    Sample height was 39 mm at RLP. 
    %7.055 g, 2.5 g/cm^3, 1.2799 cm^2 
  }
    \label{fig:finite_size}
  \end{center}
\end{figure*}

{\it Dependence on particle diameter. --} The inset of
Fig.~\ref{fig:diameter} indicates that $\phi_{\rm RLP}$ decreases with
particle diameter; however, this decrease is due to lower volume
fraction near a container wall, an effect known since the earliest
studies \cite{scott:60}.  This effect is explained in
Fig.~\ref{fig:finite_size}(a): since particles cannot penetrate the
container wall, voids are larger there and the volume fraction of the
layer adjacent to the boundary is lower than $\phi_{\rm bulk}$
measured in the core of the sample. The difference between $\phi_{\rm
apparent}$ averaged over the whole container and $\phi_{\rm bulk}$
increases with the ratio $d$/$D$ and produces the trend displayed in
the inset of Fig.~\ref{fig:diameter}.

We examined the finite size effect using X-ray tomography
\cite{sakellariou:04}, %aste:04,aste:05 as illustrated in
Fig.~\ref{fig:finite_size}. In each run positions of $1.5 \times 10^5$
spheres were measured with a resolution of better than 0.1\% of a sphere diameter
\cite{aste:07}. Figure \ref{fig:finite_size}(c) shows the difference
between the apparent volume fraction % $\phi_{\rm app}$ calculated
using all particles and the bulk volume fraction (measured using the
Voronoi volumes \cite{aste:07} of all particles that are at least $4d$
away from the container walls). For $Q<10$ ml/min, $\phi_{\rm apparent}$
values from the tomographic measurements agree with the results
(dotted curve) for the 277 $\mu$m particles in
Fig.~\ref{fig:diameter}, but for $Q > 10$ ml/min the tomographic
values are larger due to unavoidable vibrations during the recording
of the tomogram.

The inset of Fig.~\ref{fig:finite_size}(c) shows the difference
between the bulk and apparent $\phi$ as a function of 
$\phi_{\rm apparent}$. A linear fit yields
\begin{equation}
\label{eq:finite_size_corr} \phi_{\rm bulk} = \phi_{\rm apparent} + 0.122
(\phi_{\rm apparent} - 0.505).
\end{equation}
For all further experiments we used only spheres with diameter 261 or
257 $\mu$m (sample E or F), and we corrected for the effect of
finite container size using (\ref{eq:finite_size_corr}).

\begin{figure}[t]
  \begin{center}
    \includegraphics[width=8cm]{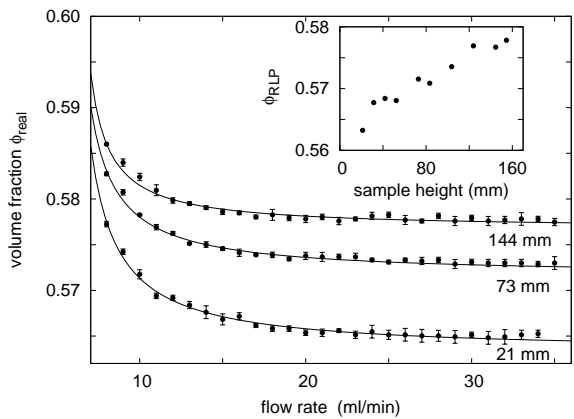}
    \caption{Dependence of $\phi$ on $Q$ for different sample heights
(261 $\mu$m spheres, sample E).  Values of $\phi$ were corrected for
finite size effect using (2). Solid lines are fits to
(\ref{eq:rlp_fit}).  Inset: the resultant $\phi_{\rm RLP}$ values as a
function of sample height.  }
    \label{fig:column_height}
  \end{center}
\end{figure}

%%%%%%%%%%%%%%%%%%%%%%%%%%%%%%%%%%%%%%%%%%%%%%%%%%%%%%%%%%%%%%%%%%%%%%%%%%%%%%%%%%%%%%%%%
{\it Influence of pressure. --} The stress inside a column of grains
differs from the hydrostatic case in two ways: (i) anisotropy -- the
horizontal stress $\sigma_{\rm xx}$ in the column differs from the
vertical stress $\sigma_{\rm zz}$. (ii) wall friction -- 
the part of the load carried by the frictional sidewalls increases with
depth $z$ below the surface. Consequently, $\sigma_{\rm xx}$ and
$\sigma_{\rm zz}$ saturate with $z$. In our analysis we use a
pressure dependence on height given by the Janssen model
\cite{sperl:06}, which assumes a constant stress ratio $K =
\sigma_{\rm xx} / \sigma_{\rm zz}$ everywhere in a sample.
Experiments show that this model is a fair approximation in the
absence of external loads \cite{vanel:99,vanel:00}.  The model gives
a saturation of pressure ($p = \sigma_{\rm zz}$) with
depth,
\begin{equation}
    \label{eq:janssen}
    p(z) = p_{\rm sat}  \left( 1 - e^{-z/l} \right),
\end{equation}
to a constant value $p_{\rm sat} = \Delta \rho g D / 4 K \mu_{\rm W}$,
where $\mu_{\rm W}$ is the coefficient of particle-wall friction, $g$
is the gravitational acceleration, $\Delta \rho$ is the density
difference between spheres and surrounding liquid, and $l = D/4 K\mu_{\rm W}$. 
Equation (\ref{eq:janssen}) indicates two ways of
controlling the pressure distribution inside the column:

I) Increasing the sample height, which increases the fraction of the
sample at $p_{\rm sat}$. If $\phi_{\rm RLP}$ increases with $p$, then
the average $\phi_{\rm RLP}$ measured by our method should increase with
sample height. This behavior is confirmed in Fig.~\ref{fig:column_height}. 

\begin{figure}[t]
  \begin{center}
    \includegraphics[angle=-90,width=8.4cm]{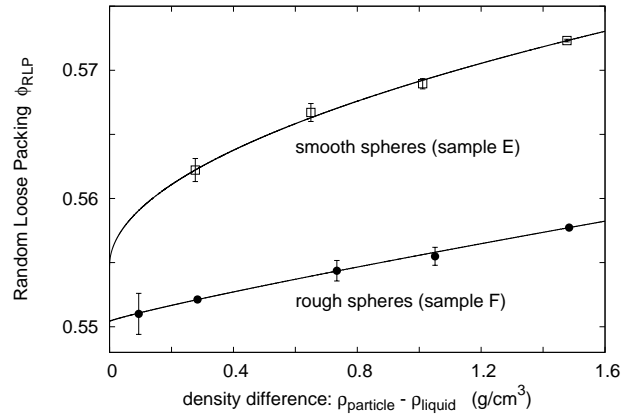}
    \caption{$\phi_{\rm RLP}$ as a function of the density difference
    between particles and fluid, for smooth (261 $\mu$m) and rough
    (257 $\mu$m) particles.  Solid lines are fits to
    (~\ref{eq:pressure_dep}).  Data have been corrected for the finite
    size effect using (\ref{eq:finite_size_corr}). Sample height was 64 mm, and the points are
    averages obtained at the highest possible flow rate, since for
    small density differences the flow rate range was too small for a
    meaningful fit with (\ref{eq:rlp_fit}).  
    % 11.2 g, 2.48 g/cm^3, 1.2799 cm^2, 0.55 
    }
    \label{fig:friction}
  \end{center}
\end{figure}

II) Decreasing the density difference $\Delta \rho$, which decreases
$p_{\rm sat}$ but keeps the pressure profile unchanged.  We increased
the fluid density to as high as 2.39 g/cm$^3$, close to the 2.48
g/cm$^3$ particle density, by adding sodium polytungstate to the
water.  Results for different $\Delta \rho$ (Fig.~\ref{fig:friction})
again confirm that $\phi_{\rm RLP}$ decreases with decreasing $p$.

A limit $\phi_{\rm RLP}^0$ would be given by matching the fluid and
particle densities, but in this limit there would be no sedimentation
and no connected granular packings would form. Therefore, we
extrapolate to determine $\phi_{\rm RLP}^0$: In the absence of theory
we follow \cite{valverde:04} and use the
pressure dependence close to the jamming point known for
frictionless static soft spheres
\cite{makse:00,ohern:02,zhang:05,majmudar:07} and frictionless
thermal hard spheres \cite{brito:06}:
\begin{equation}
\label{eq:pressure_dep} \phi_{\rm RLP} =  \phi_{\rm RLP}^0 + \left(
\frac{\Delta \rho}{a}\right)^\alpha ,
\end{equation}
where we identify $\Delta \rho \sim p_{\rm sat}$. A fit of $\phi_{\rm
RLP}$ for smooth particles (sample E) in Fig.~\ref{fig:friction}
yields $\phi_{\rm RLP}^0 = 0.555 \pm 0.006$. The value of $\alpha=
0.51 \pm 0.25$ is approximate because our derivation of
(\ref{eq:pressure_dep}) did not take into account the $\phi$
dependence of $K$~\cite{vanel:99}.

%%%%%%%%%%%%%%%%%%%%%%%%%%%%%%%%%%%%%%%%%%%%%%%%%%%%%%%%%%%%%%%%%%%%%%%%%%%%%%%%%%%%%%%%%%%%%

{\it Influence of frictional properties. --} Figure \ref{fig:friction}
shows that $\phi_{\rm RLP}$ for the rough spheres was lower than for
the smooth spheres.  For the rough spheres a fit to
(\ref{eq:pressure_dep}) yields $\phi_{\rm RLP}^0 = 0.550 \pm 0.001$
and $\alpha = 0.89 \pm0.16$.  The decrease of $\phi_{\rm RLP}$ and
$\phi_{\rm RLP}^0$ with increasing friction agrees with another
experiment \cite{menon:07}, model \cite{srebro:03},  and simulations
\cite{zhang:05,shundyak:07,silbert:08,pica_ciamarra:08}.

%%%%%%%%%%%%%%%%%%%%%%%%%%%%%%%%%%%%%%%%%%%%%%%%%%%%%%%%%%%%%%%%%%%%%%%%%%%%%%%%%%%%%%%%%%%%%%%%

{\it Discussion. --} 
Our experimental results and numerical simulations \cite{zhang:05,shundyak:07} both show that RLP is well
defined in an operational sense: in the limit of infinitesimally slow preparation, the volume fraction of
a sample converges to $\phi_{RLP}$, independent of the details of preparation. Care should be taken in
comparing theory for frictionless hard spheres with the experimental results, in part because of the
different possible ways of defining mechanical stability \cite{torquato:01,torquato:07}.

The observation of a well-defined $\phi_{RLP}$ can be considered within the framework of a statistical
mechanics of static granular material \cite{edwards:07}, 
where a configurational entropy $S$ is defined as the logarithm
of the number of mechanically stable configurations for a given $\phi$, $p$, and friction coefficient. Two
different approaches can explain RLP using two different assumptions of how $S$ depends on $\phi$. 
The first approach assumes that RLP is the smallest $\phi$ where $S$ becomes larger than zero. 
This is compatible with the existence of looser, highly ordered configurations  \cite{torquato:07}, 
as their number seems not to grow exponentially with system size, so $S = 0$.

The second approach is supported by numerical results on the number of stable configurations of
frictional discs \cite{pica_ciamarra:08}, where $S$ has a maximum at RLP.  
This idea agrees with slow sedimentation 
leading to RLP: it is simply the most probable configuration. If the sedimentation speed is increased,
the additional kinetic energy allows the system to explore the local energy landscape and find 
rarer but lower potential energy (denser) configurations. Further, 
the maximum of $S$ and therefore RLP moves to higher values of $\phi$ 
for decreasing friction \cite{pica_ciamarra:08}.
This agrees with our results and with simulations of frictionless disks that have a 
maximum of $S$ at Random Close Packing \cite{gao:06}. Our results indicate 
also that increasing $p$ shifts the maximum of $S$ in a similar way.

%%%%%%%%%%%%%%%%%%%%%%%%%%%%%%%%%%%%%%%%%%%%%%%%%%%%%%%%%%%%%%%%%%%%%%%%%%%%%%%%%%%%%%%%%%%%%%%%%%%%%%

{\it Conclusions. --} Mechanically stable packings of spheres prepared with 
increasing sedimentation time display a lower bound of their volume fraction, 
$\phi_{\rm RLP}$, which depends on the pressure and the coefficient of friction 
but not on the diameter of the spheres. In the limit of zero pressure we have found a new 
lowest value of $\phi_{\rm RLP}$, $0.550 \pm 0.001$.

{\it Acknowledgments. --} We thank Massimo Pica Ciamarra and Antonio
Coniglio for sharing their unpublished results, and Brandon McElroy
for his assistance with the Camsizer.  Further we thank
W.D.~McCormick, Narayanan Menon, Charles Radin, Leo Silbert, Jack
Swift, and participants of the 2007 Aspen Jamming workshop for
helpful discussions.  This work was supported by Robert A. Welch
Foundation Grant F-0805.

%\bibliography{rlp}

\end{document}